\documentclass[twocolumn,10pt,cleanfoot,cleanhead]{asme2e}

\usepackage{graphicx}
\usepackage{amsmath,amssymb}
\usepackage{booktabs}
\usepackage{makecell}
\usepackage{caption}
\usepackage{subcaption}
\usepackage{url}
\usepackage{xcolor}
\newcommand{\ch}[1]{{\color{black}{#1}}}
\title{Ortho2CAD: 3D CAD generation from orthographic drawings using vision language models}

\author{Aditya Joglekar \qquad Amit Regmi \qquad Kenji Shimada \qquad \textbf{Levent Burak Kara}\thanks{Address all correspondences to lkara@andrew.cmu.edu} \affiliation{ Department of Mechanical Engineering\\Carnegie Mellon University\\ Pittsburgh, PA, 15213, USA}}

\begin{document}

\maketitle    

\begin{abstract}
{\it Engineering design intent is often communicated through rasterized orthographic drawings. However, downstream workflows require editable and parametrically defined 3D computer-aided design (CAD) models. To bridge this gap, we introduce vision language model (VLM) frameworks specifically designed to translate rasterized orthographic drawings into editable CadQuery code, which can then be converted into 3D CAD models. Firstly, due to unavailability of large scale orthographic drawing datasets, we create a pythonOCC-based drawing generator that renders first-angle orthographic projections from STEP models, with dashed hidden lines and bounding box dimensions, and generate over 1 million drawings from existing 3D CAD model datasets. We also create a dataset of 100 drawings with manually dimensioned features. We show that supervised fine-tuning applied on small open-source VLMs when paired CadQuery code is available improves reconstruction accuracy on corresponding test sets. For datasets without code labels, geometry-grounded reinforcement learning is performed which uses generated-solid intersection-over-union (IoU) with ground truth solid as the reward, improving code validity and cross-dataset generalization. Then, an inference time self-refinement framework for frontier VLMs is introduced which repeatedly repairs invalid codes and compares orthographic projections of generated 3D models with the input drawing to revise the CadQuery code. Our self-refinement framework with GPT 5.5 achieves \(100\%\) valid code generation and the highest mean IoU across all test sets, with an average relative improvement of more than \(11\%\) over the next-best method. We show that leveraging VLMs can effectively pave the way forward for orthographic drawing to 3D CAD reconstruction. Our implementation is available at \url{https://github.com/AdityaJoglekar/Ortho2CAD}.}
\end{abstract}

\section{Introduction}

3D computer-aided design (CAD) models are the dominant representation for mechanical parts and are directly usable in downstream engineering workflows such as manufacturing planning, simulation and cost estimation. Yet, much of the design intent in practice is communicated through 2D engineering technical drawings, specifically multi-view orthographic drawings that provide a standardized way to encode topology and dimensions. While these drawings may be authored digitally (often in vector form), the format most frequently exchanged in real manufacturing workflows is often an image-based (raster) drawing due to ease of sharing, quality assurance and intellectual-property protection because image drawings are not editable \cite{zhang2023component}. This prevalence of raster drawings creates a major impediment to 3D reconstruction automation because unlike vector formats that provide direct scripted access to geometric and semantic entities, raster drawings typically require human inspection to extract the information. Together, these realities create a persistent disconnect between how designs are specified and communicated (often as rasterized 2D drawings) and how they must be reconstructed as editable 3D CAD by designers and engineers for subsequent applications.

Recent progress in large language models (LLMs) and vision language models (VLMs) has enabled direct generation of CAD programs from text descriptions, perspective images and point clouds exemplified by works such as \cite{alrashedy2024generating}, \cite{guan2025cad}, \cite{doris2025cad} and \cite{kolodiazhnyi2025cadrille}. They produce CadQuery code (CadQuery is a Python based parametric CAD scripting library) and evaluate the resulting solids using geometric similarity metrics such as intersection-over-union (IoU). While these approaches demonstrate the CadQuery code-generating abilities of LLMs and VLMs, they do not tackle the task of converting 2D engineering drawings to CadQuery code, despite orthographic drawings being a standard modality for part design communication and a prime input modality designers deal with for 3D CAD model creation. 

We address two challenges in the task of converting orthographic drawings to CadQuery codes (and thus 3D CAD) in this work. The first challenge is unavailability of large-scale datasets of standardized orthographic drawings paired with source CAD models, which are required for VLM training. To enable learning at scale, we create a pythonOCC-based drawing generator that renders first-angle orthographic projections with bounding box dimensions from STEP models, and generate drawings for over 1 million 3D CAD models from existing datasets. We also create a dataset of 100 drawings with manually dimensioned features. The second challenge is the availability of training supervision varies substantially across CAD datasets. We show that supervised fine-tuning (SFT) applied on small open-source VLMs when paired CadQuery code supervision is available improves reconstruction accuracy on corresponding test sets. For datasets without availability of code supervision, geometry-grounded reinforcement learning (RL) is performed which uses generated-solid intersection-over-union (IoU) with ground truth solid as the reward, improving code validity and cross-dataset generalization. Further, we compare and benchmark `the utility of frontier VLMs' with `the utility of SFT and RL applied to small open-source VLMs' on this task. We show through 4 benchmark datasets that our proposed self-refinement inference time scaling framework with the latest frontier VLMs significantly outperforms all other models on all the datasets in terms of 3D reconstruction accuracy. We briefly discuss the different components of our work next.

\textbf{Data synthesis.} Due to the unavailability of large-scale engineering drawing data, we create a scalable data generation python code using the opensource pythonOCC library to synthesize orthographic drawings with three views and critical dimension annotations from existing CAD repositories. We generate these drawings for the DeepCAD \cite{wu2021deepcad}, ZeroToCAD1m \cite{ataei2026zero} and Fusion 360 Reconstruction \cite{willis2021fusion} datasets and publish the data generation pipeline for further use by the community. Zhang et al. \cite{zhang2023automatic} release a 2D orthographic drawings dataset constructed from the Fusion 360 database and using FreeCAD \cite{FreeCAD}, but it only consists of simple 2981 samples and we could not find their exact generation pipeline on any public platform. Zhang et al. \cite{zhang2025reinforcement} use the ABC dataset \cite{koch2019abc} and FreeCAD \cite{FreeCAD} for creating close to 70k 2D orthographic drawings but the drawings do not have key dimensions and do not have dashed hidden lines, thus not following the standard conventions. We present one of the largest open-source resources of orthographic drawings, and importantly publish an automatic generation python code for orthographic drawings with dashed hidden lines and key dimensions derived from public CAD datasets, enabling systematic study of drawing-to-editable-CAD generation under controlled, reproducible conditions. In addition, we create a dataset of 100 drawings with manually dimensioned features.

\textbf{Supervised finetuning.} On DeepCAD (GenCAD-Code) and ZeroToCAD1m datasets, where ground-truth CadQuery codes/programs exist, supervised fine-tuning of Qwen3VL-8B-Instruct \cite{bai2025qwen3} is performed to translate orthographic drawings into CadQuery code. Compared to the CAD-Coder \cite{doris2025cad} baseline, fine-tuning a stronger, newer VLM backbone on the same underlying objective improves reconstruction quality. It also indicates that orthographic drawings can be an effective conditioning signal for code synthesis when paired code supervision exists.

\textbf{Reinforcement learning on Fusion 360 reconstruction dataset.} Almost all CAD datasets, including Fusion360 reconstruction, do not provide aligned ground-truth CadQuery codes in a form directly usable for supervised code synthesis. This setting motivates reinforcement learning (RL) with geometric feedback, where the model can be optimized using reward signals computed from the CAD produced by executing generated code, rather than relying on target codes. We use a Dr. GRPO \cite{liu2025understanding} style group-relative objective with sequence-level optimization and apply it to orthographic-drawing-to-3D-CAD reconstruction via CadQuery code on the Fusion360 dataset, showing that policy optimization driven by geometric similarity rewards substantially improves 3D reconstruction accuracy over supervised-only baselines and compensates for the absence of ground-truth CadQuery codes.

\textbf{Using Large Frontier Models.} Using off the shelf large frontier language and vision models for CAD script generation has been shown to achieve a strong performance \cite{alrashedy2024generating, badagabettu2024query2cad}. Engineering design is inherently iterative, and the ability of these models to generate, critique, and refine candidate programs aligns naturally with the looped feedback cycles used to converge on high‑quality parts. There has been rapid progress in the capabilities of recent large frontier models, and we showcase their utility in our engineering drawing to CadQuery code task in this work. We further show that inference time scaling with an orthographic projection image feedback generator can substantially boost reconstruction accuracy and is unmatched by the supervised finetuning and reinforcement learning frameworks for all datasets.

\begin{figure*}
\centering
\begin{subfigure}{\textwidth}
\centering
\includegraphics[width=0.85\textwidth]
{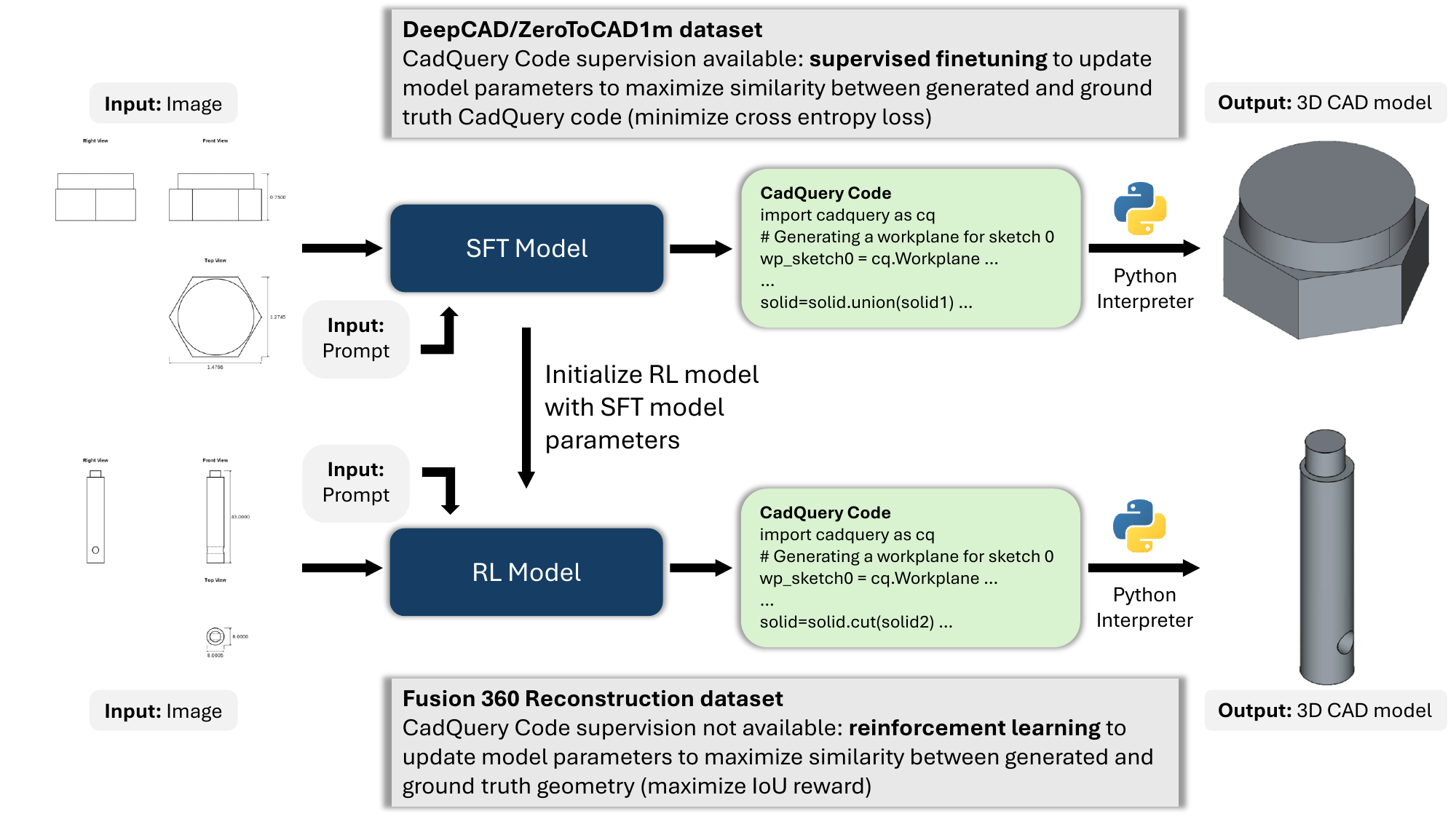}
\caption{Supervised finetuning and reinforcement learning}
\end{subfigure}
\centering
\begin{subfigure}{\textwidth}
\centering
\includegraphics[width=0.84\textwidth]
{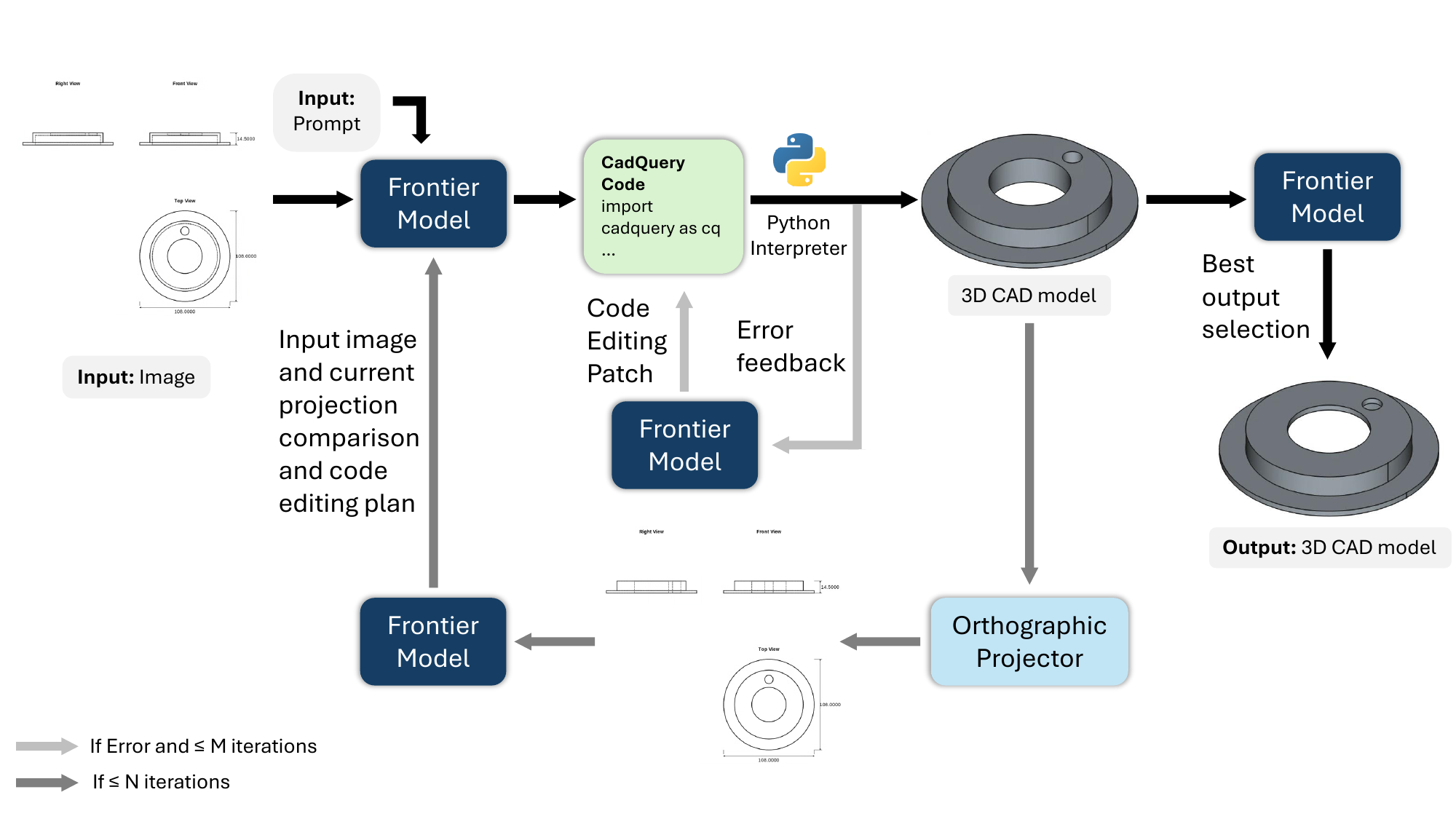}
\caption{Inference time scaling with large frontier models}
\end{subfigure}
\centering
\caption{Ortho2CAD: Our methods take an orthographic drawing image and a text prompt as input and generate a CadQuery code that compiles into a STEP model. We present three methods: supervised finetuning (SFT), reinforcement learning (RL) and inference time scaling (ITS). With open source models, we perform SFT on a pretrained VLM when CadQuery code labels are available; for datasets without supervision, we initialize from an SFT model and apply geometry-grounded RL. If closed source large frontier models are available for inference, we apply our ITS framework to get state-of-the-art results.
}
\label{fig:Ortho2CADmethod}
\end{figure*}

Overall, our approach (Figure \ref{fig:Ortho2CADmethod}) reframes editable CAD reconstruction using VLMs around an engineering native specification format of orthographic drawings. We show that using small open source models for supervised finetuning and reinforcement learning can give significant improvements in results, but if closed source large frontier models are available for inference, using an inference time scaling framework with them outperforms everything else. The key contributions are:
\begin{enumerate}
    \item a supervised VLM fine-tuning and a RL-based training framework for substantially improving the results of pretrained open source models.
    \item an inference time scaling framework with an orthographic projection image feedback generator demonstrating state-of-the-art drawing-to-CAD synthesis with large frontier models.
    \item an orthographic drawing dataset and python code for generating large-scale orthographic drawing datasets from CAD models.
\end{enumerate}

\section{Literature Review}

Learning-based geometry generation with prediction of representations such as voxels, meshes and point clouds \cite{nichol2022point,poole2022dreamfusion,siddiqui2024meshgpt,xiang2025structured} has seen significant development over the past few years. For engineering tasks, there has also been a development in predicting Boundary representation (B-rep) solids \cite{xu2024brepgen, liu2025hola}. Downstream tasks benefit from feature-level editability and several recent methods highlighted below increasingly treat CAD modeling as synthesizing codes that enable editability and which can be executed to obtain B-rep solids.

\textbf{CAD datasets and representations.}
A key enabler for CAD synthesis is the availability of large-scale CAD datasets, where having sequential operations for CAD models in the dataset can be very useful. The ABC dataset \cite{koch2019abc} is a large repository of 1 million complex CAD models. Fusion 360 Gallery dataset \cite{willis2021fusion} provides a reconstruction dataset of 8,625 designs, comprising sequential sketch and extrude modeling operations. DeepCAD \cite{wu2021deepcad}, which consists of a subset of models from the ABC dataset, introduces more than 150k CAD models paired with construction sequences and demonstrates that Transformer-style sequence models can learn distributions over CAD programs. CAD-Coder \cite{doris2025cad} extends this dataset and creates the GenCAD-Code dataset consisting of CAD-model image and CAD Query code pairs. 

\textbf{LLM-based CAD program synthesis.}
An end-to-end transformer-based autoregressive network is proposed by \cite{khan2024text2cad} to generate parametric CAD models from input texts. Guan et al. \cite{guan2025cad} reformulate text-to-CAD as the generation of CadQuery scripts. This representation enables a seamless integration with existing LLMs. Their approach proposes a two stage pipeline of  supervised fine-tuning and then reinforcement learning with a geometric and format reward. Govindarajan et al. \cite{govindarajan2025cadmium} generate a dataset annotated with high-quality, human-like descriptions and fine-tune code-LLMs to generate CAD sequences represented in a JSON-based format from natural language descriptions.
However, all the above approaches do not consider the problem of image to CAD generation.

\textbf{VLM-based CAD program synthesis.}
Moving beyond text, VLM-based methods have begun to infer CAD programs from images to support reverse engineering and design-from-visual-input workflows. Several works showcase utility of large frontier VLMs for this task. Query2CAD \cite{badagabettu2024query2cad} takes a language prompt as input and then creates a FreeCAD macro with a LLM which is refined using renders and a caption generating model. Alrashedy et al. \cite{alrashedy2024generating} introduce CADCodeVerify which takes a language prompt as input and iteratively verifies and improves 3D objects generated from VLM generated CadQuery code by firstly rendering the object and then producing feedback from the VLM and also deviation correction prompts for the VLM. MEDA \cite{panta2025meda} improves CADCodeVerify's results with a multi-agent framework. Seek-CAD \cite{li2025seek} takes a textual description of a part (language prompt) as input and utilizes DeepSeek-R1 \cite{guo2025deepseek} locally for their proposed SSR CAD script generation and uses rendered perspective views and chain-of-thought feedback from an accompanying VLM for refinement.
Image2CADSeq \cite{li2025image2cadseq} creates a pipeline to reverse engineer CAD models by processing images as input and generating CAD sequences. However, it limits itself to a small transformer model and a simplified domain specific language (DSL) and does not utilize pretrained VLMs.
Doris et al. \cite {doris2025cad} firstly create the GenCAD-Code dataset and release an open-source VLM (CAD-Coder) trained to produce CadQuery codes from images, demonstrating that finetuning open source VLMs such as LLaVA \cite{liu2023visual} can generate editable CAD. 
The majority of VLM-based program synthesis work focuses on perspective renders, product images or language prompts and do not consider engineering drawings as the specification modality.

\textbf{Reinforcement learning and execution-grounded optimization.}
Reinforcement learning algorithms such as GRPO \cite{shao2024deepseekmath,guo2025deepseek} and its evolutions \cite{yu2025dapo, liu2025understanding} have recently gained substantial traction as core post-training methods for LLMs/VLMs. These outcome-grounded approaches move beyond static supervised fine-tuning by optimizing policies against explicit reward signals derived from human preferences or programmable/verifiable objectives, enabling tighter control over generation quality and alignment. 
Cadrille \cite{kolodiazhnyi2025cadrille} proposes a multimodal CAD reconstruction framework that uses RL to improve executable program generation with feedback computed from executed CAD outputs, showing that RL can substantially improve reconstruction quality beyond supervised initialization. 
Relatedly, \cite{niu2025intent} propose a multimodal chain-of-thought guided RL framework for CAD code generation. However, both the above works do not consider the problem of orthographic drawings as input. Also, since we were unable to locate any publicly available implementation of their RL code at the time of writing of this paper, we exclude these methods in our results comparisons. RL has also been investigated specifically for orthographic drawing inputs in \cite{zhang2025reinforcement}. This paper achieves high quality reconstructions of parametric CAD from 2D orthographic drawings using a DQN-style RL agent, but it assumes cleaned, fully-informative three-view inputs and omits common drafting conventions like dashed hidden lines and key dimensions.
Because it depends on strict preprocessing and idealized drawing assumptions, the approach can be hard to scale to messy real-world drawings.
Unlike VLM based frameworks, it does not leverage pretrained VLM generalization or a portable program representation like CadQuery code for broader scalability. 

Taken together, the existing literature establishes (i) LLM/VLM code synthesis as a scalable path to editable CAD generation and (ii) engineering drawings, specifically multi-view orthographic projections, remain underrepresented as an input modality in the recent CadQuery-focused VLM literature despite being the dominant format for communicating manufacturing-ready design intent. This motivates treating orthographic drawings as an input for executable CAD code generation using VLMs and our work on dataset generation, SFT, RL and inference time scaling (self-refinement) with large frontier models. We present a brief comparison of the different types of VLMs and their usage in this work in Table \ref{tab:vlm_categories}.

\begin{table*}[t]
\centering
\small
\begin{tabular}{lcc}
\toprule
\makecell[l]{\textbf{VLM}} &
\makecell[l]{\textbf{Example}} &
\makecell[l]{\textbf{Notes}} \\
\midrule
\makecell[l]{Small open\\ source model} &
\makecell[l]{Qwen3-VL-8B-Instruct} &
\makecell[l]{Off-the-shelf model performance is not great.\\ Can perform SFT/RL/inference locally.\\ 4 H100 GPUs used in this work.} \\
\midrule
\makecell[l]{Large open\\ source model} &
\makecell[l]{Qwen3-VL-235B-A22B-Instruct} &
\makecell[l]{Off-the-shelf model performance is good.\\ Requires several large GPUs for\\ performing SFT/RL/inference locally.\\ Not done in this work due to insufficient compute\\ availability. Inference can be done using API calls\\ but not explored in this work.} \\
\midrule
\makecell[l]{Frontier closed\\ source model} &
\makecell[l]{GPT 5.5} &
\makecell[l]{Off-the-shelf model performance is\\ the best possible. Inference can be done using\\ API calls, which is explored in this work.} \\
\bottomrule
\end{tabular}
\caption{Comparison of the different types of VLMs and their usage in this work.}
\label{tab:vlm_categories}
\end{table*}

\section{Method}
We present details of our dataset generation procedure and our frameworks here.
\subsection{Dataset Generation}
\label{sec:orthodatasetgen}
Large-scale datasets of dimensioned, multi-view orthographic engineering drawings aligned with 3D CAD are not readily available, which limits systematic development and evaluation of drawing-to-editable-CAD models. To address this gap, we create a scalable data generation python code that converts 3D CAD shapes into standardized orthographic drawings using pythonOCC, an open-source Python wrapper for OpenCascade that is straightforward to install and run in a lightweight virtual environment.

Given an input CAD model in STEP format, our python code loads the B-Rep geometry and renders three standard orthographic views (front, top, and right) using first angle projection and following technical drawing conventions with solid strokes for visible edges, dashed strokes for hidden edges and three key dimensions shown. These dimensions define the bounding box of the part, but do not capture all the geometric features pertinent to the drawing. For example, the inner diameter of a hollow cylinder will not be present and will have to be deciphered from the ratio of the inner diameter and a key dimension available in that direction. However, we believe that including some dimensions, unlike most prior works that do not include any dimensions, is an important step towards perfectly dimensioned drawings, and automation of dimensioning all features in a model is left for future work. Our drawings resemble the raster artifacts used in industrial workflows and preserve geometric fidelity to the original CAD.

We generate orthographic drawings with the STEP files in the DeepCAD, ZeroToCAD1m and Fusion 360 reconstruction datasets. We also create a Fusion 360 reconstruction manual dimensions dataset, where we use the same test subset from the Fusion 360 reconstruction dataset and use FreeCAD \cite{FreeCAD} for manually creating and placing dimensions for the features in the part for a more realistic dataset compared to datasets with just the bounding box dimensions. We present a comparison of the datasets in Table \ref{tab:datasetscomp}. DeepCAD and Fusion 360 reconstruction datasets have only sketch and extrude operations and have human authored models, while the ZeroToCAD1m dataset has several complex operations and has synthetically generated models. For the DeepCAD and ZeroToCAD1m datasets, we set timeouts of 10 and 60 seconds respectively (ZeroToCAD1m has more complex drawings) for generation of a drawing, which serves the purpose of faster generation and also mimicking real scenarios where sometimes certain views may be missing or the drawing may be incomplete.

The data generation code can be applied to any available dataset that contains STEP files.

\begin{table*}[t]
\small
\centering
\begin{tabular}{ccccc}
\toprule
\makecell[c]{\textbf{Criterion}} &
\makecell[c]{\textbf{DeepCAD (with} \\ \textbf{GenCAD-Code)}} &
\makecell[c]{\textbf{ZeroToCAD1m}} &
\makecell[c]{\textbf{Fusion 360} \\ \textbf{Reconstruction}} &
\makecell[c]{\textbf{Fusion 360} \\ \textbf{Reconstruction}\\
\textbf{Manual}\\
\textbf{Dimensions}}\\
\midrule
\makecell[c]{Number of\\ Examples} &
\makecell[c]{GenCAD-Code \cite{doris2025cad}\\ train/test/val\\
split of\\ 147289/7355/9027;\\ 100 test subset\\ for final\\ evaluation same\\ as \cite{doris2025cad}} &
\makecell[c]{Official\\ train/test/val\\
split with successful\\ orthographic\\ projections\\ 977000/9977/9972;\\ 100 test subset for\\ final evaluation} &
\makecell[c]{Official\\ train/test split\\ of 6900/1725;\\ 100 test subset\\ for final evaluation} &
\makecell[c]{Only 100 test\\ examples for final\\ evaluation with same\\ ground truth STEP\\ files as Fusion 360\\ Reconstruction 100\\ test subset.}\\
\midrule
\makecell[c]{Ground Truth\\ STEP files source} &
\makecell[c]{Wu et al.\cite{wu2021deepcad}} &
\makecell[c]{Ataei et al.\cite{ataei2026zero}} &
\makecell[c]{Willis et al. \cite{willis2021fusion}} &
\makecell[c]{Willis et al. \cite{willis2021fusion}}\\
\midrule
\makecell[c]{Usable for SFT\\ (Ground truth\\ CadQuery code\\ available)} &
\makecell[c]{Yes, used\\ in this work} &
\makecell[c]{Yes, used\\ in this work} &
\makecell[c]{No} &
\makecell[c]{No}\\
\midrule
\makecell[c]{Usable for RL} &
\makecell[c]{Yes, but not\\ explored in\\ this work} &
\makecell[c]{Yes, but not\\ explored in\\ this work} &
\makecell[c]{Yes, used\\ in this work} &
\makecell[c]{No} \\
\midrule
\makecell[c]{Usable for ITS} &
\makecell[c]{Yes, used\\ in this work} &
\makecell[c]{Yes, used\\ in this work} &
\makecell[c]{Yes, used\\ in this work} &
\makecell[c]{Yes, used\\ in this work}\\
\bottomrule
\end{tabular}
\caption{
Comparison of DeepCAD, ZeroToCAD1m, Fusion 360 Reconstruction and Fusion 360 Reconstruction Manual Dimensions datasets. The orthographic projection images corresponding to each of the models in these datasets are created in this work.
}
\label{tab:datasetscomp}
\end{table*}

\subsection{Supervised Finetuning}
\label{sec:sft}

\paragraph{Problem setup.}
In the supervised regime, our goal is to learn a conditional distribution over executable CadQuery codes given an input query.
Let $q$ denote the input query consisting of an orthographic drawing image and an instruction prompt, and let $\tau = (y_1,\dots,y_T)$ denote the target CadQuery token sequence.
We fine-tune a pretrained vision-language model (VLM) $\pi_{\theta}$ to maximize the likelihood of the ground-truth CadQuery code.
We use Qwen3-VL-8B-Instruct \cite{bai2025qwen3} as the pretrained backbone due to its reasonable parameter count relative to the CAD-Coder baseline ($\sim$13B parameters) while still providing a strong multimodal foundation. We train separate models on the DeepCAD (GenCAD-Code) and ZeroToCAD1m datasets. This is because they have very different CadQuery code styles, including different final variable names for exporting the STEP files, and we believe it would be better to keep it simple first rather than mix the two datasets directly. We keep the mixed training framework for future work.

\paragraph{Input/output formatting.}
Each training sample consists of: (i) a three-view dimensioned orthographic drawing $I$ (PNG raster), and (ii) a fixed natural-language instruction prompt
``Generate the CADQuery code needed to create the CAD for the provided image. Just the code, no other words.''
which matches the prompt template used by CAD-Coder. While this prompt can be improved, we choose to use this for simplicity and plan to improve on this prompt in future work. For off-the-shelf models where finetuning is not performed, we used a similar strategy as \cite{doris2025cad} and appended to the prompt: ``Assign the final solid to the variable ‘solid’ in the last line of code. Do not export or visualize the solid.'' to match the evaluation of generated code.
The supervision target is the corresponding CadQuery code $\tau$, which can be executed to produce a 3D CAD solid.

\paragraph{Objective.}
We minimize the standard autoregressive cross-entropy (negative log-likelihood) loss over the CadQuery token sequence conditioned on the query $q=(I,\text{prompt})$:
\begin{equation}
\mathcal{L}_{\mathrm{SFT}}(\theta)
=
-\mathbb{E}_{(q,\tau)\sim \mathcal{D}}
\left[
\sum_{t=1}^{T} \log \pi_{\theta}(y_t \mid q, y_{<t})
\right]
\label{eq:sft_nll}
\end{equation}
where $\mathcal{D}$ is our dataset of generated orthographic drawings from either the DeepCAD dataset or the ZeroToCAD1m dataset and corresponding CadQuery code pairs.
This objective directly encourages token-level correctness of the generated CadQuery code.

\paragraph{Training details.}
We fine-tune Qwen3VL-8B-Instruct for $5$ epochs on each dataset separately (refer to Table \ref{tab:datasetscomp} for train/test/val splits). We follow the standard recommended configuration for Qwen3VL for finetuning: the vision model fixed is kept fixed, and the multimodal MLP/projection layers that map vision features into the LLM token space and the language backbone (LLM) are kept trainable.
We use a learning rate of 1e-5, per-device batch size of $4$, and $4$ gradient accumulation steps, which we found to provide stable optimization in this setting. The Qwen3VL-8B-Instruct training on DeepCAD dataset takes 11 hours on $4$ H100 GPUs. For the CAD-Coder baseline (which is supervised finetuning of the LLaVA model), we use their recommended settings, and the training on \ch{our DeepCAD-derived orthographic drawing
dataset paired with GenCAD-Code CadQuery codes} also takes 11 hours on $4$ H100 GPUs. For the ZeroToCAD1m training of Qwen3VL-8B-Instruct, it takes 75 hours on $4$ H100 GPUs. We do not train the LLaVA architecture on this dataset to save computation as we see significant improvement over finetuned LLaVA by the finetuned Qwen3VL-8B-Instruct model in results for the DeepCAD dataset.

\subsection{Reinforcement Learning}
\label{sec:rl}

\paragraph{Motivation: learning without code supervision.}
Many practical CAD reconstruction datasets provide ground-truth geometry (e.g., STEP) but do not provide aligned, standardized CadQuery codes for supervised code synthesis.
To train in this setting, we use execution-grounded reinforcement learning (RL): the VLM generates CadQuery code and the code is executed to produce a CAD solid.
Since code execution and solid generation are not differentiable operations, RL provides a natural mechanism to improve the policy (VLM) using scalar feedback computed from the executed geometry.

\begin{figure}
\centering
\includegraphics[width=0.45\textwidth]{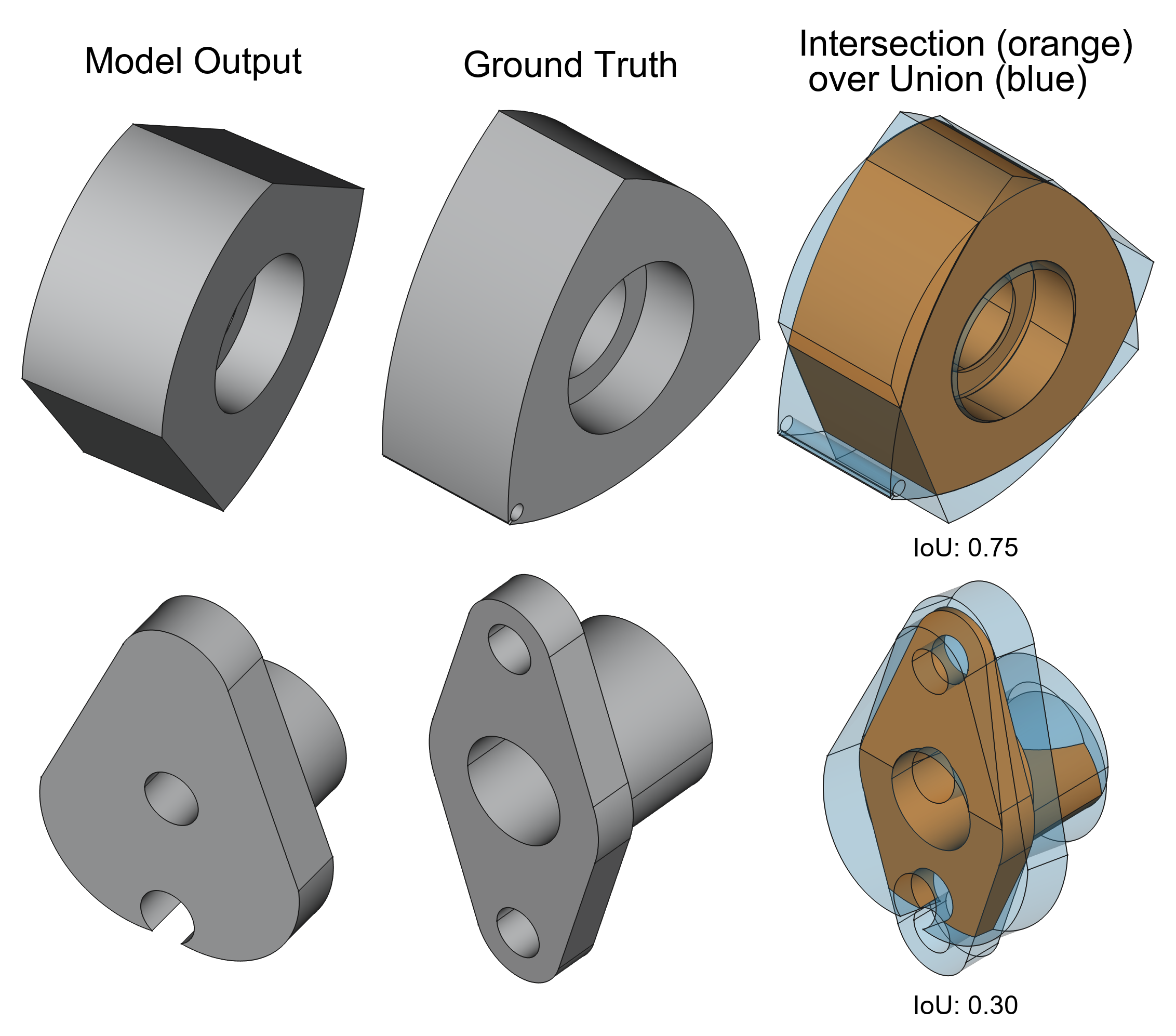}
\caption{Visualization of IoU. The model output and ground truth are normalized and aligned \cite{doris2025cad} for calculating the IoU.
}
\label{fig:iouvis}
\end{figure}

\paragraph{Reward from executable geometry.}
For a query $q=(I,\text{prompt})$, let $\tau$ denote a generated CadQuery code and $S(\tau)$ denote the solid obtained by executing $\tau$ (exported to a STEP file for IoU computation).
Let $S^{\star}(q)$ denote the ground-truth solid (STEP file).
We define the reward as the solid intersection-over-union:
\begin{equation}
R(q,\tau) \;=\; \mathrm{IoU}\left(S(\tau),\, S^\star(q)\right)
\label{eq:reward_iou}
\end{equation}
where $\mathrm{IoU}$ is defined as in \cite{doris2025cad}. In Figure \ref{fig:iouvis}, we show visualization of the IoU for two parts. If the generated code is invalid (parse/runtime failure or no valid solid), we set $R(q,\tau)=0$ and do not apply any additional penalty term. 

\paragraph{Why Dr. GRPO with sequence-level optimization.}
We draw on the design principles of Dr. GRPO \cite{liu2025understanding}. It analyzes how group-based reinforcement learning objectives can introduce length and normalization induced biases and proposes simple modifications that yield more reliable optimization dynamics when learning is driven by sparse, terminal outcome rewards.
This closely matches our setting: the reward is only observed after executing a complete CadQuery program, so overly long or unstable generations increase execution-failure rates and effectively censor the learning signal, making it important to avoid update rules that implicitly favor longer (but not better) outputs.
Concretely, we adopt mean-centered group advantages without additional length or variance normalization, and we follow the GSPO perspective \cite{zheng2025group} that when rewards are assigned at the sequence level, optimization should also be performed at the sequence level (via full-trajectory log-likelihoods) rather than via token-level credit heuristics, which improves stability for long completions and better matches the granularity of executable-geometry rewards.

\paragraph{Group sampling and advantages.}
At each iteration, for each query $q$ we sample a group of $G$ candidate codes
$\{\tau^{(g)}\}_{g=1}^{G}$ from the current policy $\pi_{\theta}(\cdot \mid q)$ and compute rewards
$R(q,\tau^{(g)})$ using Eq.~\ref{eq:reward_iou}.
We compute a mean-centered group advantage (with no additional reward scaling):
\begin{equation}
A^{(g)}(q)
=
R(q,\tau^{(g)}) - \frac{1}{G}\sum_{j=1}^{G} R(q,\tau^{(j)}).
\label{eq:advantage_mean}
\end{equation}
This produces a purely relative learning signal: for a given drawing, above-average samples are reinforced
and below-average samples are suppressed.
We perform one gradient update using each sampled group and then discard it; new groups are sampled
from the updated policy in the next iteration.

\paragraph{Autoregressive sequence likelihood.}
An autoregressive model factorizes the probability of a code
$\tau^{(g)}=(y^{(g)}_1,\dots,y^{(g)}_{T^{(g)}})$ into a product of conditional token probabilities:
\begin{equation}
\pi_{\theta}(\tau^{(g)}\mid q)
=
\prod_{t=1}^{T^{(g)}} \pi_{\theta}\!\left(y^{(g)}_{t}\mid q, y^{(g)}_{<t}\right).
\label{eq:ar_factorization}
\end{equation}
Taking the logarithm converts this product into a sum:
\begin{equation}
\log \pi_{\theta}(\tau^{(g)}\mid q)
=
\sum_{t=1}^{T^{(g)}} \log \pi_{\theta}\!\left(y^{(g)}_{t}\mid q, y^{(g)}_{<t}\right).
\label{eq:ar_loglik}
\end{equation}

\paragraph{Policy optimization objective.}
We update the policy using a sequence-level objective that increases the likelihood of sampled codes
with higher relative reward within each group:
\begin{equation}
\mathcal{L}_{\mathrm{RL}}(\theta)
=
-\mathbb{E}_{q\sim\mathcal{D}}
\left[
\frac{1}{LG}\sum_{g=1}^{G} \log \pi_{\theta}(\tau^{(g)}\mid q)\,A^{(g)}(q)
\right]
\label{eq:l_grpo_onpolicy}
\end{equation}
where $L$ is a fixed global scaling constant: the maximum completion length.
Intuitively, Eq.~\ref{eq:l_grpo_onpolicy} increases the probability of complete CadQuery codes that achieve higher IoU than other samples drawn for the same input drawing, without requiring any ground-truth CadQuery codes.

\paragraph{Training details.}
We use the Qwen3VL-8B-Instruct finetuned on the DeepCAD dataset from section \ref{sec:sft} and train on the Fusion 360 reconstruction dataset using the dataset provided train/test split \cite{willis2021fusion}. We choose the DeepCAD SFT model and not the ZeroToCAD1m SFT model as a starting point as the DeepCAD SFT model has a higher output code validity and IoU for the Fusion 360 reconstruction dataset and this dataset has only sketch and extrude operations similar to the DeepCAD dataset. We keep the vision model fixed during training, and the multimodal MLP/projection layers and the language backbone (LLM) are kept trainable. We use the same prompt format as in SFT for the respective models. Each training sample provides an orthographic drawing (generated from the ground-truth STEP via our pipeline) and the ground-truth STEP geometry used to compute the IoU reward.
No ground-truth CadQuery codes are used during RL.
We provide a summary of the hyperparameters used in the RL training in Table \ref{tab:rlhyperparams}.
\begin{table}[t]
\centering
\begin{tabular}{lc}
\toprule
\makecell[l]{\textbf{Hyperparameter}} &
\makecell[c]{\textbf{Value}} \\
\midrule
\makecell[l]{Number of Epochs} &
\makecell[c]{2}\\
\midrule
\makecell[l]{Learning Rate} &
\makecell[c]{1e-6}\\
\midrule
\makecell[l]{Effective batch size} &
\makecell[c]{64}\\
\midrule
\makecell[l]{Number of generations ($G$) per query $q$} &
\makecell[c]{8}\\
\bottomrule
\end{tabular}
\caption{
Reinforcement learning hyperparameters
}
\label{tab:rlhyperparams}
\end{table}
A maximum model length of 8192 similar to Qwen3VL official settings is used in SFT, and thus we use a maximum completion length ($L$) of 7680 tokens in RL (we obtain this by subtracting from 8192 the input tokens count and a small buffer for change in input). The training takes 100 hours on 4 H100 GPUs.

\subsection{Inference Time Scaling}
\paragraph{Motivation.} While SFT with small open sourced models may give good results on in distribution data, its out of distribution results are not expected to be usable. Also, RL on the small open source models may help in cases where a few STEP files are available for training, but again it is not expected to work great on any other out of distribution data. Moreover, simple SFT and RL reduces the ability of these already small and not highly capable VLMs to generate and ingest feedback. Closed-source large frontier language and vision-language models, which have been trained using massive resources, have advanced rapidly and exhibit strong performance on CAD script generation, even without task-specific training. Their ability to generate, evaluate, and refine candidate designs is well aligned with the inherently iterative nature of engineering design, where feedback is repeatedly incorporated to improve part quality. While cost per token is involved and some concerns for giving proprietary input images may be there, they avoid the overhead of supervised fine-tuning or reinforcement learning: no GPU setup, dataset curation, or training pipeline is required, and the interface is a simple API call. We test these large frontier models and also create a simple self refinement (inference time scaling) framework with these models which is modular and easy to setup and access.
\paragraph{Self refinement framework.}
We use the same input formatting as defined in section \ref{sec:sft} for the image and prompt for off-the-shelf models. With these inputs, the VLM outputs the first pass CadQuery code. Now, an automatic check is done for code validity and whether a STEP file can be exported. If not, the VLM is asked to output an editing patch that can replace the existing CadQuery code. The VLM is given instructions for correct output format so that the output patch can be utilized for editing. The edited code is again checked for validity and STEP export, and the editing loop is continued for until the CadQuery code successfully exports a STEP file, or if M editing iterations are reached (we have kept M=5 for the outputs in the results section). Once we have the STEP file, we use our orthographic drawing generator (section \ref{sec:orthodatasetgen}) to get the orthographic projections corresponding to the model in the STEP file. We ask the VLM to compare the input image and the current projection image and to output a plan for any changes required to the CadQuery code for making the corresponding model's projections closer to the input image. We also give the accumulated conversation history to the VLM in this VLM call. With this plan, and again all the accumulated conversation history, the VLM is called to output a better CadQuery code. The process of orthographic drawing generation and feedback and new CadQuery code is done for N iterations (we have kept N =2 in the outputs in the results section, i.e. 2 loops of feedback and new code generation are completed).

It may not always be the case that the CadQuery code of the last iteration is the best. Since we do not have the ground truth 3D CAD model during the testing phase, we choose to use the input image again as a guide to compare the different CadQuery codes output in different iterations. We call the VLM for one last time, with the accumulated conversation history, and ask it to compare each orthographic projection image corresponding to each CadQuery code with the input image and to score them, and select the best orthographic projection image, and thus the best CadQuery code and STEP model. We use 8192 as maximum new output tokens for consistency with the other pretrained open source, SFT and RL models. 

\subsection{Evaluation details}
Following \cite{doris2025cad}, we evaluate geometric reconstruction accuracy using the intersection-over-union (IoU) between the solid produced by executing the generated CadQuery code (exported to STEP) and the ground-truth CAD solid (ground-truth STEP). For model outputs that cannot be executed, we assign them IoU $=0$. This helps avoid inflating the mean IoUs of VLMs that have a large number of invalid outputs. We report IoU on a subset of $100$ test examples for each dataset due to the high computational cost and time of generating and evaluating the output of all VLMs. For the DeepCAD dataset, we use the same subset as \cite{doris2025cad}, while for the ZeroToCAD1m and Fusion 360 Reconstruction datasets we randomly sample 100 examples from the official test split, and for the Fusion 360 Reconstruction Manual Dimensions dataset we use the same 100 example IDs as the Fusion 360 Reconstruction Dataset.

\section{Results}

This section presents a quantitative and qualitative evaluation of the proposed approaches.

We consider three baseline families that collectively span (i) open-source pretrained vision-language models, (ii) a domain-specialized model trained specifically for image-to-CadQuery generation, and (iii) closed-source frontier models representative of the best off-the-shelf capability at the time of writing:
\begin{enumerate}
\item \textbf{Qwen3VL family (open-source pretrained VLMs).}
We evaluate pretrained models from the Qwen3VL family to establish a strong, reproducible open-source reference point for multimodal CadQuery code synthesis. These models are attractive because they provide modern VLM capabilities (image understanding + code generation) while remaining directly finetunable and deployable with limited compute, making them a practical baseline for follow-on supervised and RL-based adaptation.

\item \textbf{CAD-Coder (domain-specialized).}
CAD-Coder is included because it is the state-of-the-art open-source model for image to CadQuery code generation and is trained explicitly for executable CadQuery synthesis. As such, it provides the most relevant specialized comparison point for our setting, allowing us to measure whether conditioning on orthographic drawings and using a stronger backbone with targeted finetuning yields improvements over the best existing CadQuery-focused VLM SFT baseline method.

\item \textbf{GPT 5.5 and Claude Opus 4.8 (closed-source, general-purpose frontier VLMs).}
We include these as representative high-capability closed-source frontier models, which at the time of writing are the latest general available models. These baselines serve to provide an estimate of how far one can go without task-specific training.
\end{enumerate}

\begin{figure*}[htbp]
    \centering

    \begin{subfigure}[b]{0.48\textwidth}
        \centering
        \includegraphics[width=\linewidth]{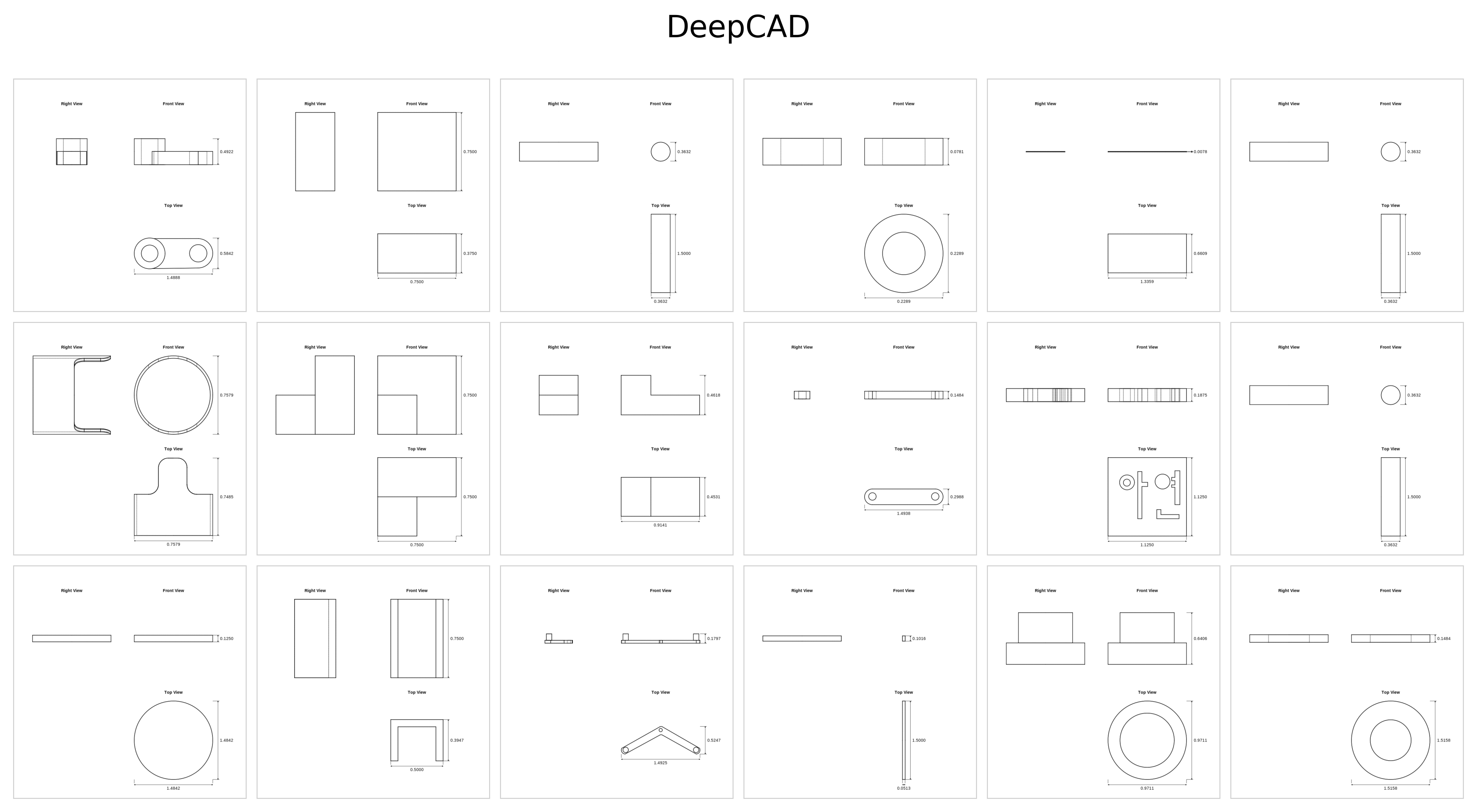}
        \label{fig:ortho2cad_inpimage_sub1}
    \end{subfigure}
    \begin{subfigure}[b]{0.48\textwidth}
        \centering
        \includegraphics[width=\linewidth]{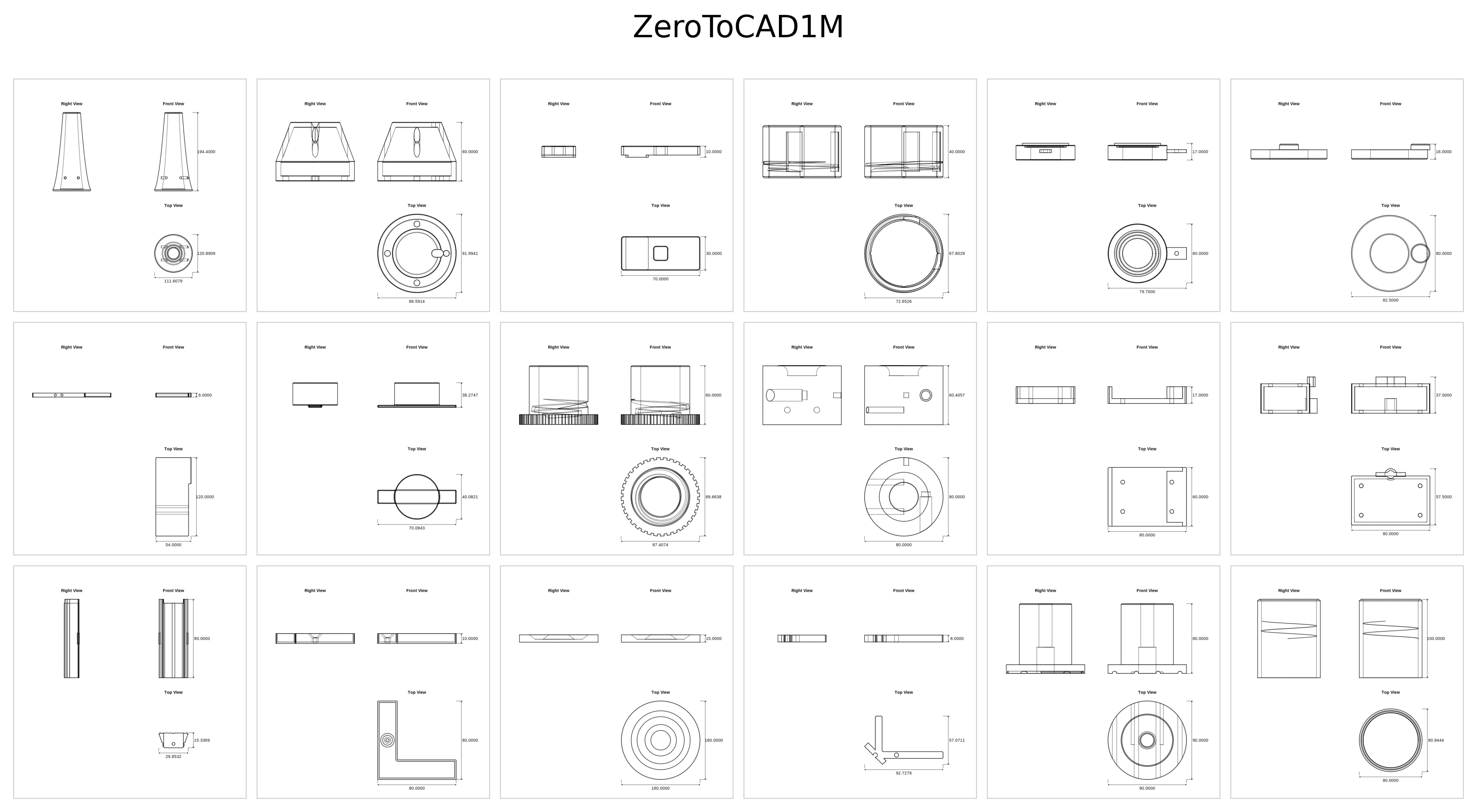}
        \label{fig:ortho2cad_inpimage_sub2}
    \end{subfigure}
    \begin{subfigure}[b]{0.48\textwidth}
        \centering
        \includegraphics[width=\linewidth]{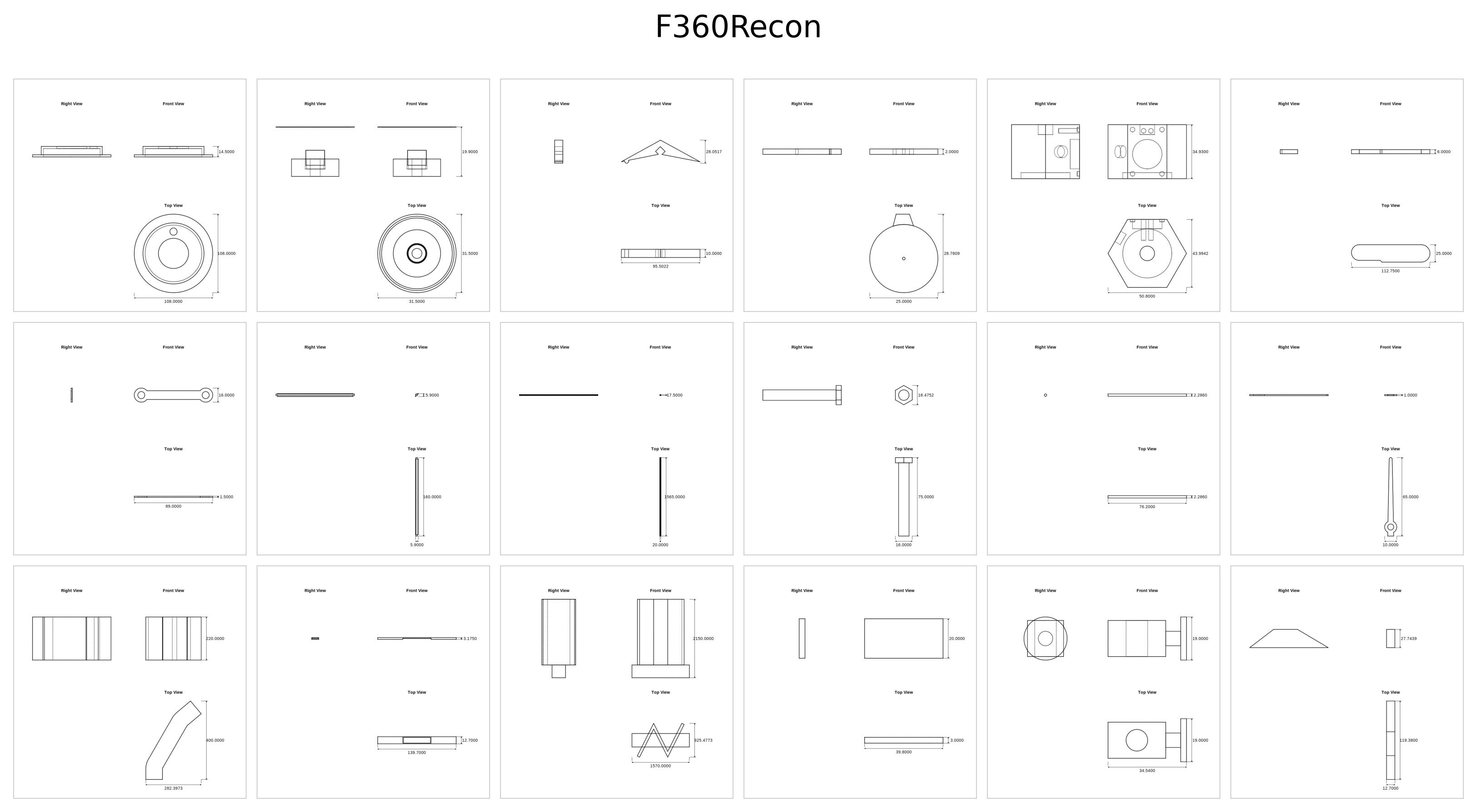}
        \label{fig:ortho2cad_inpimage_sub3}
    \end{subfigure}
    \begin{subfigure}[b]{0.48\textwidth}
        \centering
        \includegraphics[width=\linewidth]{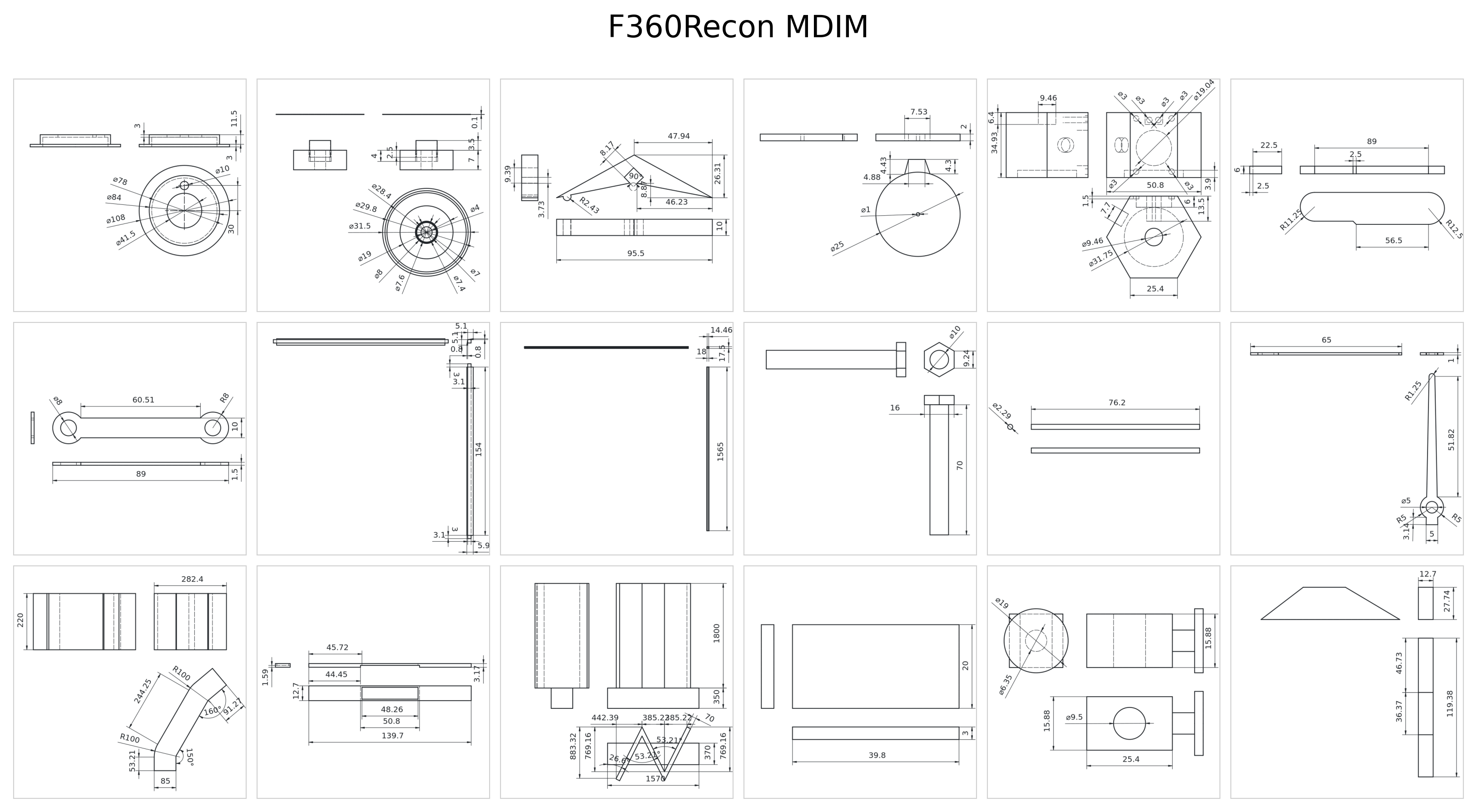}
        \label{fig:ortho2cad_inpimage_sub4}
    \end{subfigure}

    \caption{Randomly sampled input images from the four test datasets. The example IDs for the F360Recon MDIM are kept the same as F360Recon.}
    \label{fig:ortho2cad_inpimage}
\end{figure*}

\begin{figure}
\centering
\includegraphics[width=0.45\textwidth]{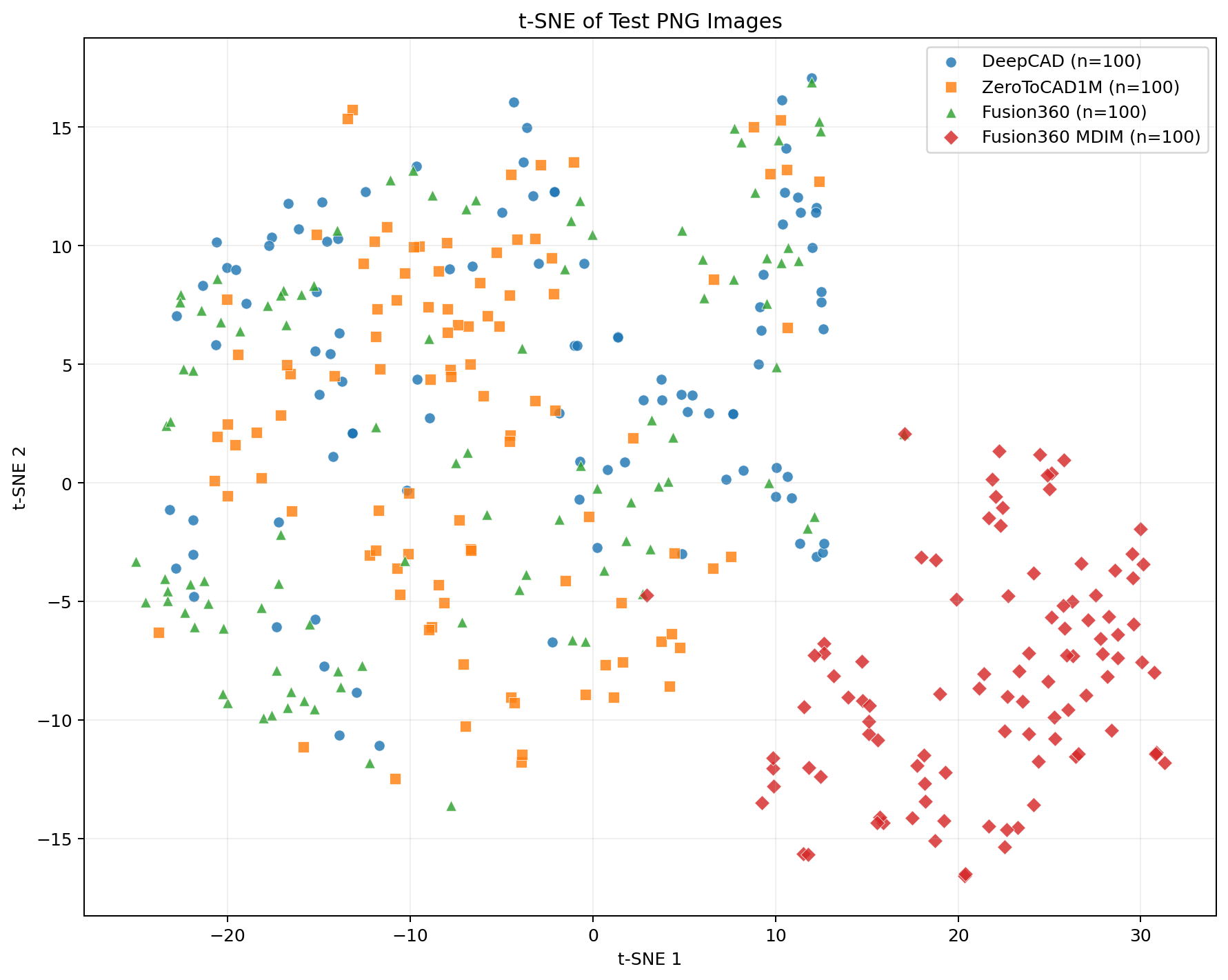}
\caption{A t-SNE plot of all the test dataset input images (4 datasets, 100 examples in each).
}
\label{fig:ortho2cad_tsne}
\end{figure}

We show a few randomly sampled input images from the four datasets in Figure \ref{fig:ortho2cad_inpimage}. We also show a t-SNE plot created with all the 400 input images in the test sets in Figure \ref{fig:ortho2cad_tsne}. We can see that the manually dimensioned images differ in style from the other test set images.

\begin{table*}[t]
\centering
\small
\setlength{\tabcolsep}{4pt}
\renewcommand{\arraystretch}{1.15}
\resizebox{\textwidth}{!}{%
\begin{tabular}{lcccccccc}
\toprule
\makecell{\textbf{Model}} &
\multicolumn{2}{c}{\makecell{\textbf{DeepCAD}}} &
\multicolumn{2}{c}{\makecell{\textbf{ZeroToCAD1m}}} &
\multicolumn{2}{c}{\makecell{\textbf{F360Recon}}} &
\multicolumn{2}{c}{\makecell{\textbf{F360Recon} \\ \textbf{Manual Dim}}} \\
\cmidrule(lr){2-3} \cmidrule(lr){4-5} \cmidrule(lr){6-7} \cmidrule(lr){8-9}
& \makecell{\textbf{Valid \%}} & \textbf{IoU}
& \makecell{\textbf{Valid \%}} & \textbf{IoU}
& \makecell{\textbf{Valid \%}} & \textbf{IoU}
& \makecell{\textbf{Valid \%}} & \textbf{IoU} \\
\midrule

\multicolumn{9}{l}{\textit{Small open-source models}} \\
Qwen3VL 8B Instruct
  & 70 & 0.3747 & 58 & 0.2043 & 58 & 0.2730 & 51 & 0.2099 \\
Qwen3VL 32B Instruct
  & 76 & 0.4919 & 55 & 0.2165 & 66 & 0.3600 & 50 & 0.3264 \\

\addlinespace[2pt]
\multicolumn{9}{l}{\textit{SFT/RL with small open-source models}} \\
CAD-Coder DeepCAD SFT
  & 100 & 0.7361 & 85 & 0.2405 & 90 & 0.2473 & 68 & 0.1657 \\
Ortho2CAD DeepCAD SFT (ours)
  & 100 & \underline{0.7922} & 95 & 0.3499 & 96 & 0.3697 & 87 & 0.1932 \\
Ortho2CAD ZeroToCAD1m SFT (ours)
  & 43 & 0.1317 & 85 & \underline{0.6167} & 65 & 0.1732 & 61 & 0.1312 \\
Ortho2CAD RL (ours)
  & 100 & 0.7761 & 100 & 0.5167 & 100 & 0.5601 & 100 & 0.2997 \\

\addlinespace[2pt]
\multicolumn{9}{l}{\textit{Large frontier closed-source models}} \\
Claude Opus 4.8
  & 99 & 0.7040 & 89 & 0.4857 & 96 & 0.6265 & 95 & 0.5785 \\
Claude Opus 4.8 ITS (ours)
  & 100 & 0.7272 & 99 & 0.5397 & 100 & 0.6625 & 100 & 0.6096 \\
GPT 5.5
  & 96 & 0.7792 & 88 & 0.5869 & 96 & \underline{0.7354} & 93 & \underline{0.6756} \\
GPT 5.5 ITS (ours)
  & 100 & \textbf{0.8458} & 100 & \textbf{0.7258} & 100 & \textbf{0.8165} & 100 & \textbf{0.7495} \\

\bottomrule
\end{tabular}%
}
\caption{A quantitative comparison of our models and the baselines across the four datasets. Best mean IoU values are shown in bold and second-best mean IoU values are underlined. For invalid outputs, we consider IoU = 0.0 when calculating the mean.}
\label{tab:ortho2cad_model_dataset_results}
\end{table*}

\begin{figure*}[htbp]
    \centering

    \begin{subfigure}[b]{0.48\textwidth}
        \centering
        \includegraphics[width=\linewidth]{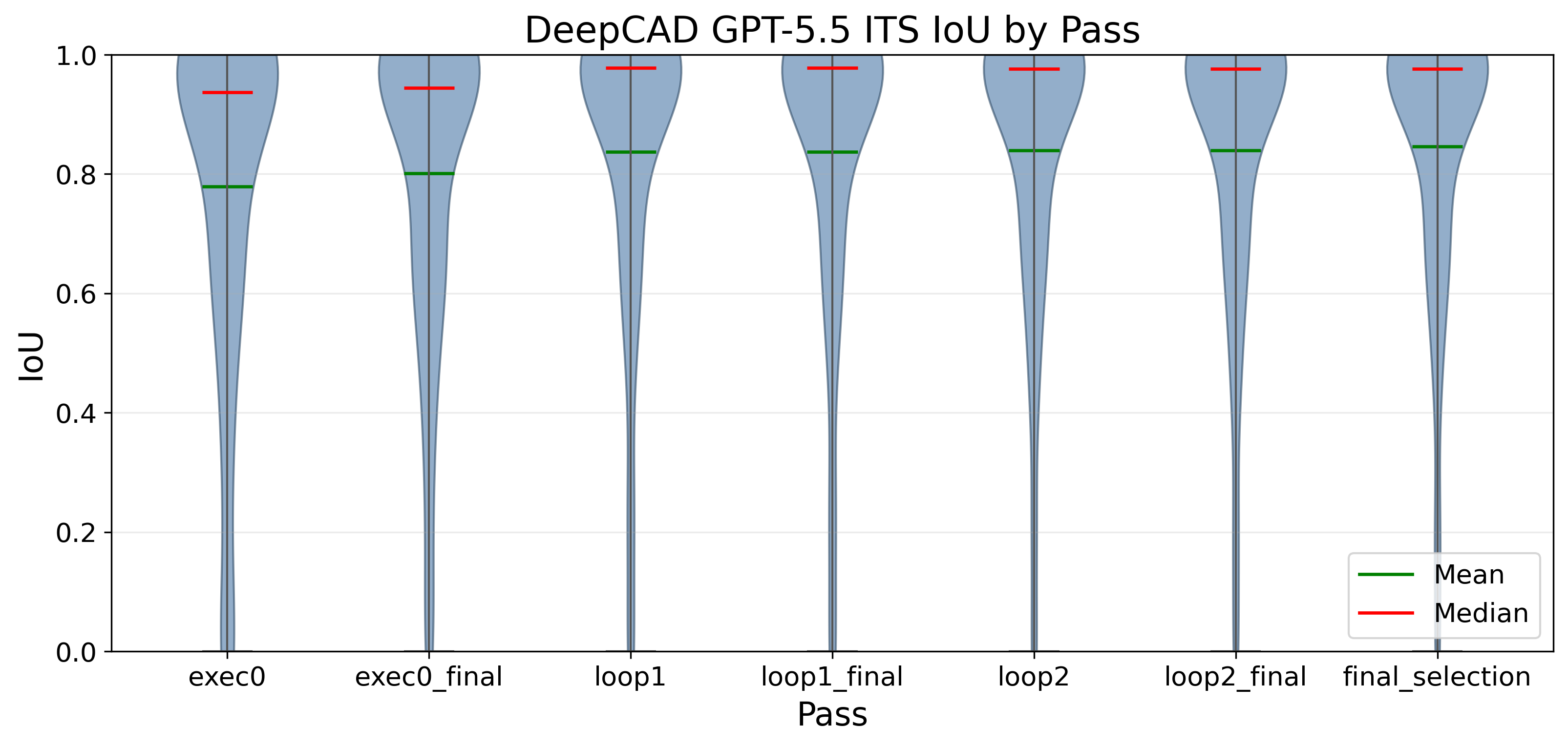}
        \label{fig:gpt5.5_its_sub1}
    \end{subfigure}
    \begin{subfigure}[b]{0.48\textwidth}
        \centering
        \includegraphics[width=\linewidth]{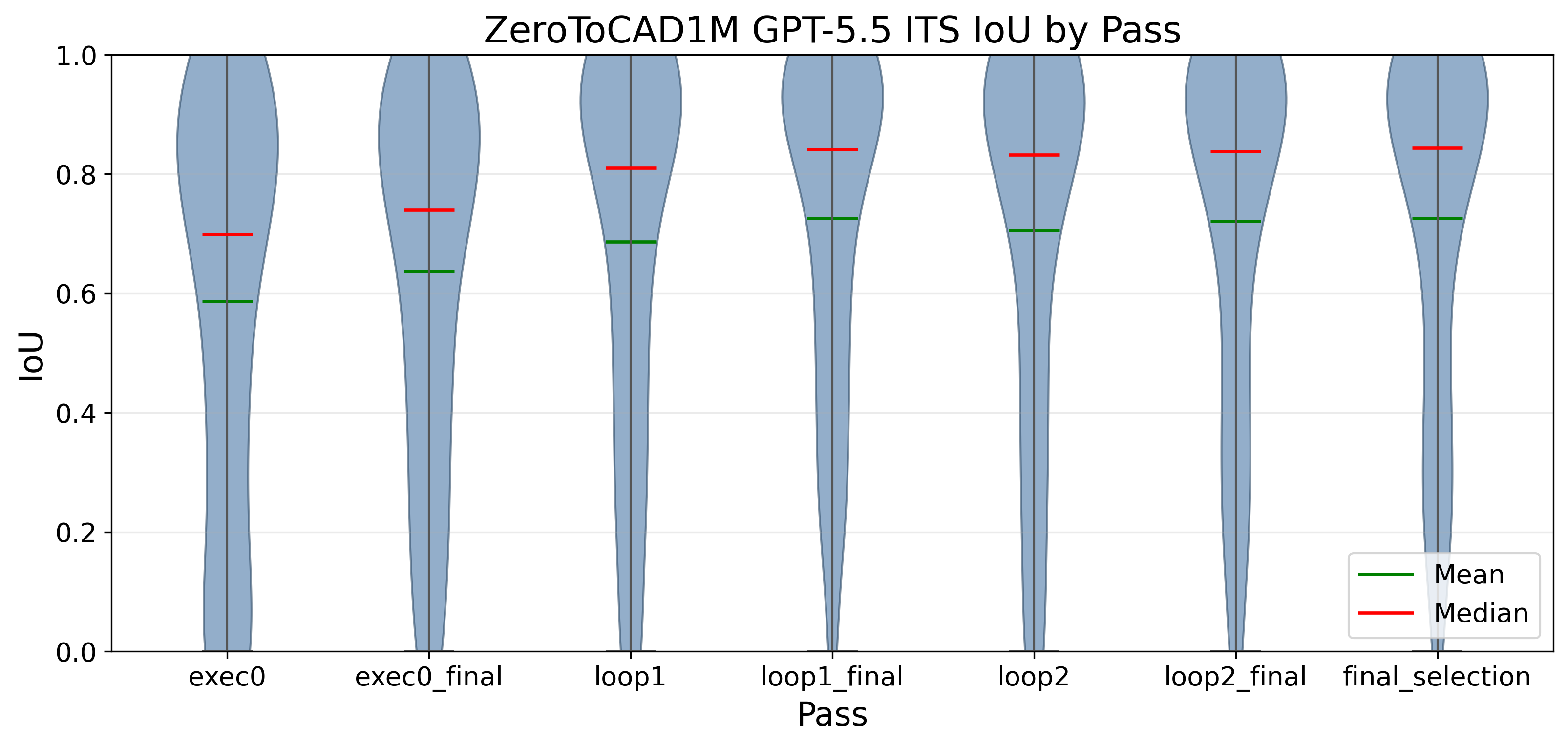}
        \label{fig:gpt5.5_its_sub2}
    \end{subfigure}
    \begin{subfigure}[b]{0.48\textwidth}
        \centering
        \includegraphics[width=\linewidth]{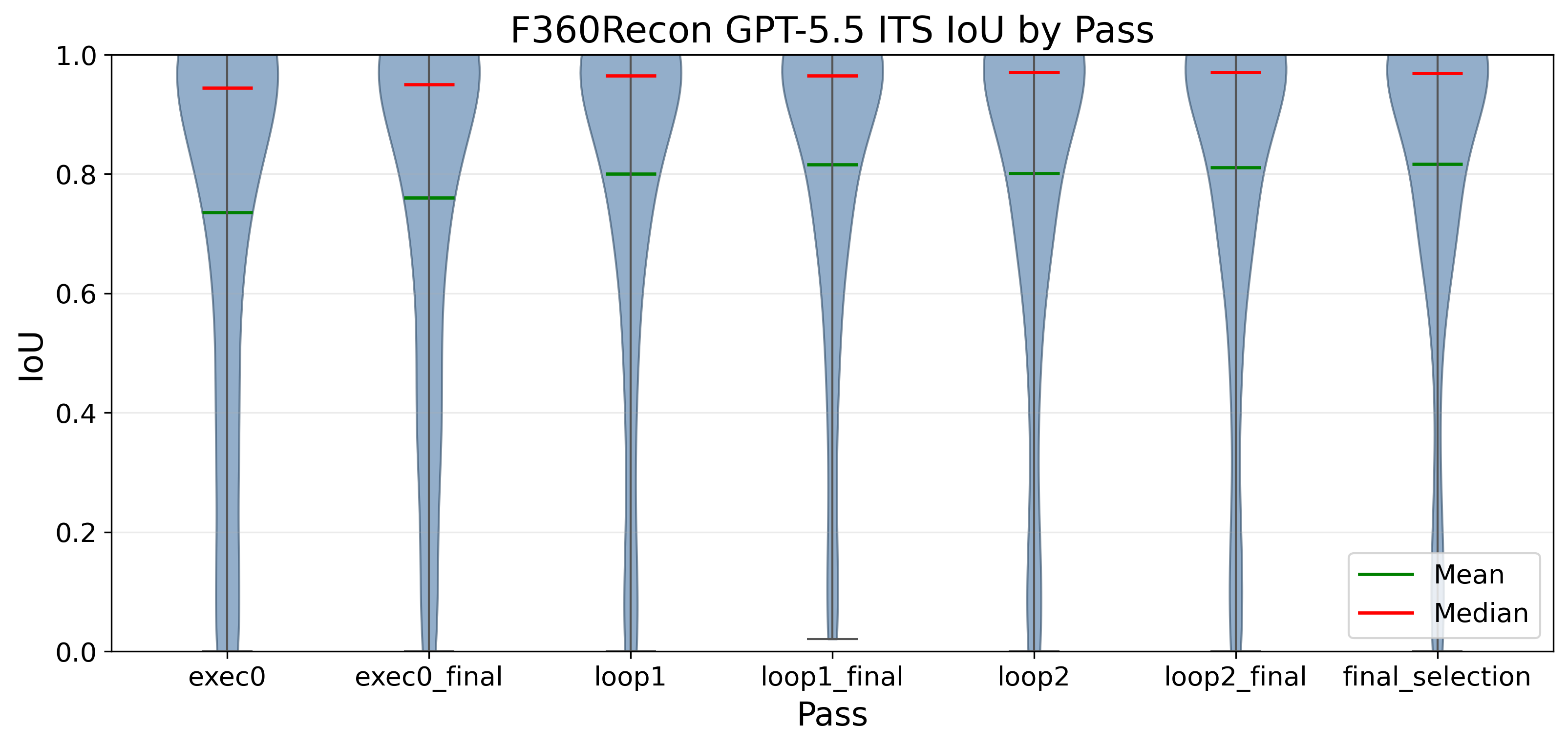}
        \label{fig:gpt5.5_its_sub3}
    \end{subfigure}
    \begin{subfigure}[b]{0.48\textwidth}
        \centering
        \includegraphics[width=\linewidth]{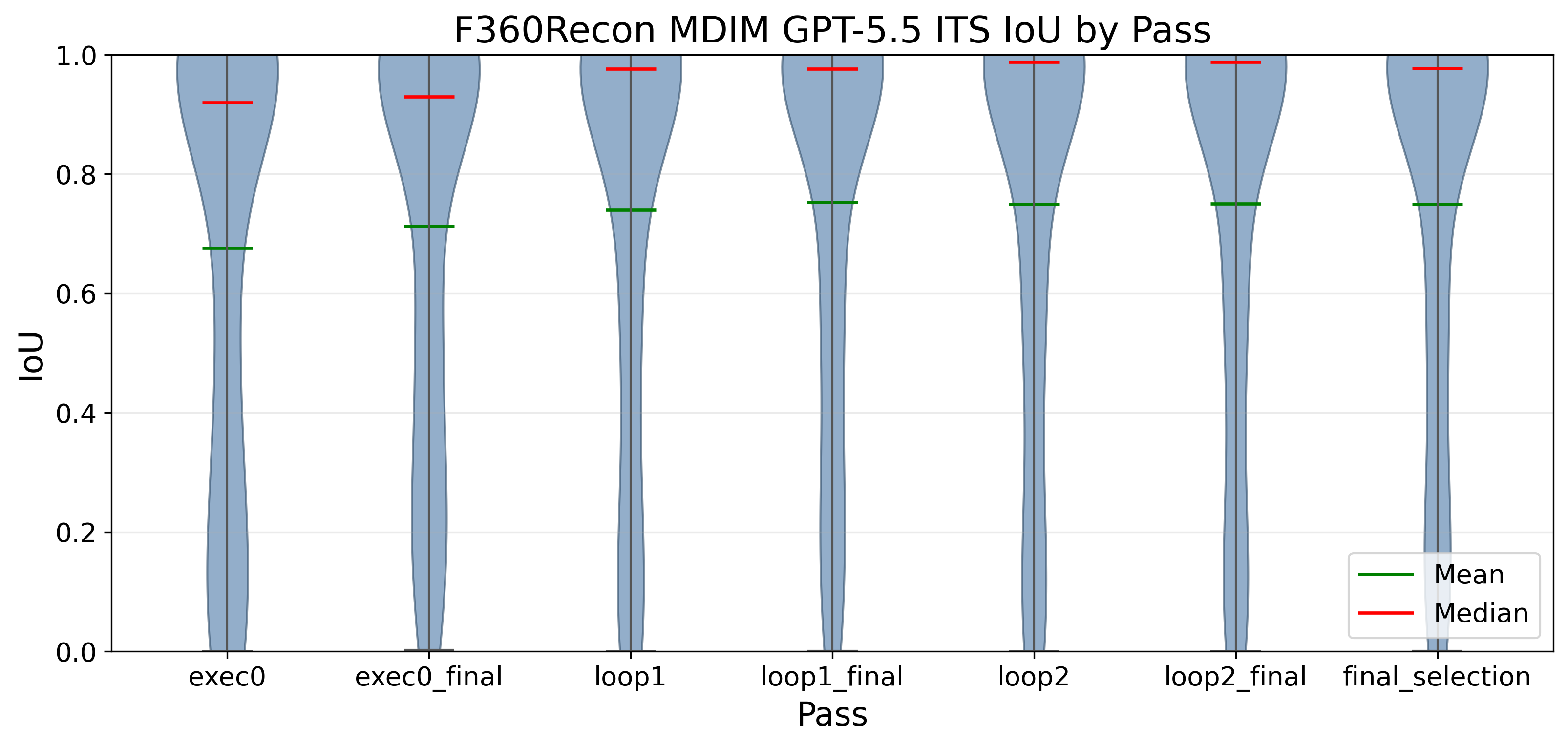}
        \label{fig:gpt5.5_its_sub4}
    \end{subfigure}

    \caption{The IoU distributions for the four test datasets with 100 examples each as the passes progress in the GPT 5.5 ITS framework. For invalid outputs, we consider IoU = 0.0. `exec0' is the GPT 5.5 result. Each pass with a `final' is the result after error correction on previous pass. Each `loop' is after image feedback and new output. The `final selection' is the model selected after image comparisons to try to get the best output from all the outputs.}
    \label{fig:gpt5.5_its_violin_plots}
\end{figure*}

\begin{figure*}
\centering
\includegraphics[width=0.9\textwidth]{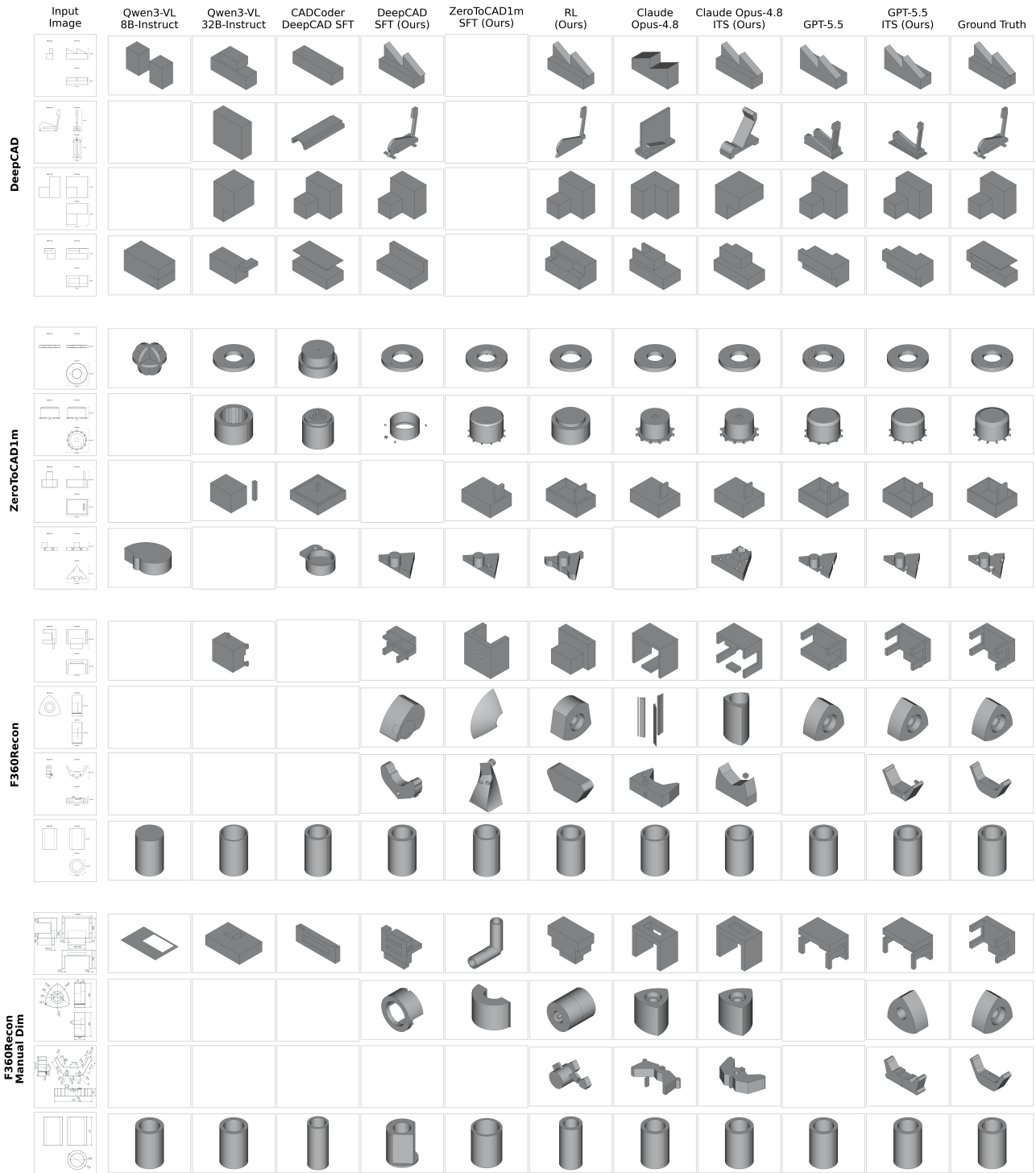}
\caption{A qualitative comparison of our models and the baselines across different examples from the four datasets. Invalid CadQuery code generation is indicated by empty cells. Our GPT 5.5 ITS model outperforms all other models.
}
\label{fig:ortho2cad_results_figure}
\end{figure*}

\begin{figure*}
\centering
\includegraphics[width=0.95\textwidth]{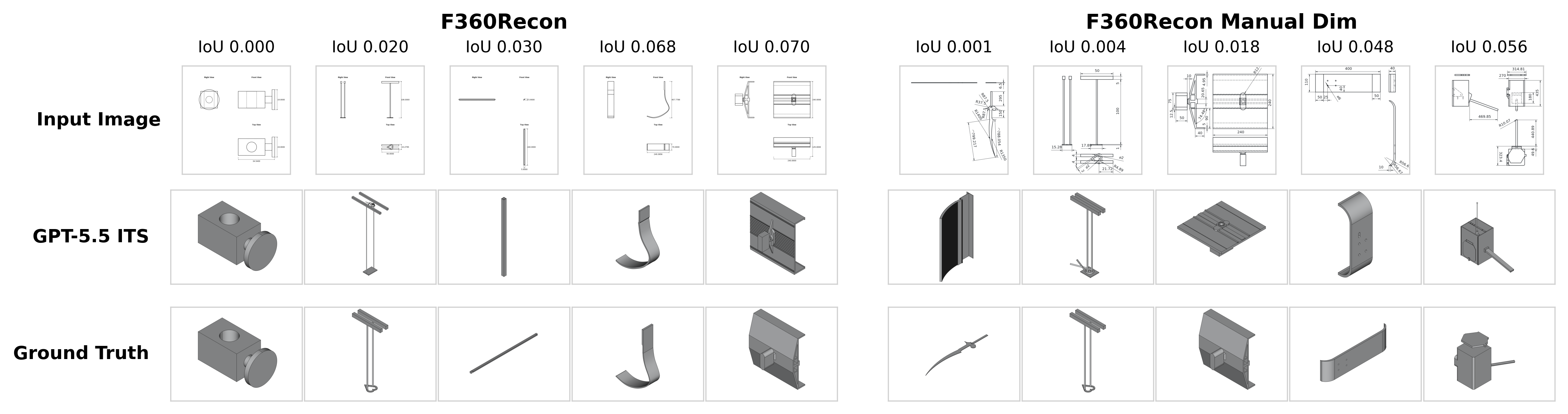}
\caption{A qualitative comparison of the worst five examples in terms of IoU for our GPT 5.5 ITS model from the F360Recon and the F360Recon Manual Dimensions datasets. We can observe that most of the examples that the VLM finds difficult to reconstruct have thin and complicated features, where the differences in the features are difficult to decipher from the input image. We also see that the first example from the F360Recon dataset has been assigned IoU as 0.0, although the reconstruction looks very accurate. This is because of a failed intersection operation in the IoU code. We plan to improve on the IoU code in future work for making it more robust.
}
\label{fig:ortho2cad_worst_iou}
\end{figure*}

\subsection{Quantitative analysis}

We report (i) \emph{Valid Codes}, the fraction of generations that are syntactically valid and execute to a solid, and (ii) \emph{IoU}, the mean intersection-over-union between the solid produced by executing generated CadQuery code and the ground-truth STEP geometry.

Table \ref{tab:ortho2cad_model_dataset_results} presents quantitative results that compare our models with the baselines.

\subsubsection{Code Validity}
The small open source models output a large number of invalid CadQuery codes, and SFT on the DeepCAD dataset significantly improves valid code generation. SFT on the ZeroToCAD1m dataset (which has complex and several different commands) improves the code validity on the ZeroToCAD1m test dataset by a large margin, however the improvement of code validity using this model on other test datasets is not that significant. For reinforcement learning, the ability to generate executable codes is very important for fast and stable training and hence we choose the DeepCAD SFT model as our starting point for RL training (see section \ref{sec:rl} for training details). RL itself clearly helps improve valid code generation as shown by our RL model which achieves 100\% valid code generation on all the four test datasets. Using the frontier models the code validity is high, and with our ITS framework it further improves, with our GPT 5.5 ITS model achieving 100\% valid code generation on all the four test datasets. 
\subsubsection{Geometric reconstruction quality}
We observe that SFT gives very high IoU for only the corresponding test dateset, i.e. the DeepCAD SFT model work great and gives the second highest IoU on the DeepCAD test set, while the ZeroToCAD1m SFT model gives the second highest IoU on the ZeroToCAD1m test set. The fall off in IoU for these models when tested on the other test sets is large, particularly for the ZeroToCAD1m SFT model.

For the RL model, since the DeepCAD SFT model was used in its initialization for the RL training, it performs well on it. The F360Recon dataset was used for the RL training, and its great performance on it demonstrates that executable-geometry feedback can effectively adapt a model to new data distributions without requiring ground-truth CadQuery codes. Notably, it also performs well on the ZeroToCAD1m test set, which it has never seen, thus highlighting that the RL training helped the model become more generalizable.

The frontier models perform consistently well across all the datasets, however it should be noted that the ZeroToCAD1m test set is the one which they find hardest. All models give worse results for the F360Recon Manual Dim dataset compared to the F360Recon dataset. The SFT and RL model performances drop off considerably on the manual dimension dataset, showcasing that changes in input image style can impact the output quite a lot. For the pretrained open source and frontier models, the slight drop off in performance for this manual dimension dataset compared to the automatic bounding box dimensions dataset could be attributed to absence of view headings and the large number of extra lines in terms of dimensions introduced in the image, which may sometimes cause the model to misinterpret the dimension lines as projection lines.

If we look at the overall results across the datasets from the frontier models, Claude Opus 4.8 ITS is inferior to the SFT models on the corresponding SFT test set and better on the other test sets, while GPT 5.5 ITS is consistently the best VLM framework for all datasets. GPT 5.5 ITS has an average relative improvement of 11.6\% across the datasets over the second best model. Also, our ITS framework gives a significant improvement in the IoU compared to using only one pass of the frontier models, thus confirming its effectiveness in this task of orthographic drawing to CadQuery code generation. We show a more detailed look at the IoU distributions after each pass in Figure \ref{fig:gpt5.5_its_violin_plots}. Maximum improvement happens after the first loop, and overall we get an average relative improvement in mean IoU over the first pass (GPT 5.5) of 13.5\% across the datasets.

\subsection{Qualitative Analysis}
In Figure \ref{fig:ortho2cad_results_figure},  we present qualitative results that compare our models with the baselines. Four representative examples from each dataset are selected.

Off the shelf small open source models do not perform well at all. We again observe here that the SFT models perform well on the test dataset corresponding to their training dataset. We also observe for many examples how GPT 5.5 ITS improves on the GPT 5.5 model pass. In the first two examples of the ZeroToCAD1m dataset, GPT 5.5 gave larger holes and thicker teeth on the cylinder, respectively, while GPT 5.5 ITS reduced the size of the holes and gave thinner teeth on the cylinder, moving closer to the ground truth.

In Figure \ref{fig:ortho2cad_worst_iou}, we showcase the examples that GPT 5.5 ITS found difficult to reconstruct from the F360Recon and F360Recon Manual Dimensions datasets.

Overall from the qualitative results, we see that for usability and utility of the produced 3D model, using GPT 5.5 ITS is necessary as most of the other VLMs are not reliable across the datasets.

\section{Conclusion, Limitations and Future Work}
The results demonstrate three key findings:
\begin{enumerate}
    \item SFT with small open source models give excellent results on the corresponding test dataset, however the results do not scale well to other test datasets.
    \item Reinforcement learning enables learning on geometric feedback, and without CadQuery program supervision, and further improves on the SFT results, however still falls short when the input image style is different for test examples compared to the training examples.
    \item Inference time scaling with frontier models gives the best results across all the datasets, with an average relative improvement of over 11\% compared directly against the next best performing model.
\end{enumerate}

There are several limitations in our current models. Firstly, the evaluation can be improved by including metrics other than IoU. Chamfer distance can be added and metrics that capture the similarities in the 2D projections of the results can also be added. Creating metrics that focus on the features and CadQuery code will be helpful. For example, differences in the inner radius of a hollow cylinder are very easy to overcome by editing the feature and hence the metric score can be high here, but if there are extra features present or absent in the result that are difficult to edit/add/remove then the metric score can be low.

Another area of potential improvement is using more complex SFT and RL training methods or creating new training methods that can improve the results. When large quantities of data and compute are available, performing SFT and RL on large open source models can give excellent results and may be useful in some cases. However, we believe reaching the performance of our frontier models ITS framework on all the datasets simultaneously will be a very difficult task using domain-specific fine-tuning or reinforcement learning strategies. Recent remarks and discussions by industry practitioners working on adjacent problems reveal strikingly similar observations \cite{zachdive}.

Overall, we find that large scale frontier models that have been trained extensively on massive amounts of data including CAD related elements, offer great performance with simple API calls, and their feedback generation and ingestion capabilities are just the things needed for designing better 3D CAD models. While our GPT 5.5 ITS model supersedes all other models, the ITS framework could still be further improved using different and better prompts and model selection criteria, and more feedback loops. Another area of exploration for future work is generating CAD scripts and instead of just an orthographic drawing generator, connecting with CAD kernels to help get even better feedback in the self refinement process by utilizing any other extra features available from these such as topological correctness and feature timelines for precise editing. Creating an orthographic projection generator with automatic feature dimensions can help improve the quality of feedback further and is another great direction for future work.

\section*{Acknowledgment} 
This work was supported by the MiSUMi Corporation. The research was conducted on ORCHARD, a high-performance cloud computing cluster made available by Carnegie Mellon University. FreeCAD software was used for visualizing the 3D parts in all the figures. We thank Faez Ahmed, Christopher McComb, Joe Joseph, Soji Yamakawa and Vedant Puri for insightful discussions.

\section*{LLM Usage}
LLMs assisted with grammar and readability changes and with boilerplate code in our experimental scripts.

\bibliographystyle{asmems4}

\end{document}